\documentstyle[11pt,epsf]{article}
\def\bea{\begin{eqnarray}}
\def\eea{\end{eqnarray}}

\def\beq{\begin{equation}}
\def\eeq{\end{equation}}
\def\ba{\beq\new\begin{array}{c}}
\def\ea{\end{array}\eeq}
\def\be{\ba}
\def\ee{\ea}

\makeatletter
\newdimen\normalarrayskip 
\newdimen\minarrayskip 
\normalarrayskip\baselineskip
\minarrayskip\jot
\newif\ifold \oldtrue \def\new{\oldfalse}
\def\arraymode{\ifold\relax\else\displaystyle\fi} 
\def\eqnumphantom{\phantom{(\theequation)}} 
\def\@arrayskip{\ifold\baselineskip\z@\lineskip\z@
\else
\baselineskip\minarrayskip\lineskip2\minarrayskip\fi}
\def\@arrayclassz{\ifcase \@lastchclass \@acolampacol \or
\@ampacol \or \or \or \@addamp \or
\@acolampacol \or \@firstampfalse \@acol \fi
\edef\@preamble{\@preamble
\ifcase \@chnum
\hfil$\relax\arraymode\@sharp$\hfil
\or $\relax\arraymode\@sharp$\hfil
\or \hfil$\relax\arraymode\@sharp$\fi}}
\def\@array[#1]#2{\setbox\@arstrutbox=\hbox{\vrule
height\arraystretch \ht\strutbox
depth\arraystretch \dp\strutbox
width\z@}\@mkpream{#2}\edef\@preamble{\halign
\noexpand\@halignto
\bgroup \tabskip\z@ \@arstrut \@preamble \tabskip\z@ \cr}%
\let\@startpbox\@@startpbox \let\@endpbox\@@endpbox
\if #1t\vtop \else \if#1b\vbox \else \vcenter \fi\fi
\bgroup \let\par\relax
\let\@sharp##\let\protect\relax
\@arrayskip\@preamble}
\def\eqnarray{\stepcounter{equation}%
\let\@currentlabel=\theequation
\global\@eqnswtrue
\global\@eqcnt\z@
\tabskip\@centering
\let\\=\@eqncr
$$%
\halign to \displaywidth\bgroup
\eqnumphantom\@eqnsel\hskip\@centering
$\displaystyle \tabskip\z@ {##}$%
\global\@eqcnt\@ne \hskip 2\arraycolsep
$\displaystyle\arraymode{##}$\hfil
\global\@eqcnt\tw@ \hskip 2\arraycolsep
$\displaystyle\tabskip\z@{##}$\hfil
\tabskip\@centering
&{##}\tabskip\z@\cr}
\begingroup\ifx\undefined\newsymbol \else\def\input#1 {\endgroup}\fi

\begin{document}

\setcounter{footnote}{1}
\def\thefootnote{\fnsymbol{footnote}}
\begin{center}
\hfill ITEP/TH-44/00\\
\hfill hep-th/0008199\\
\vspace{0.3in}
{\Large\bf On Noncommutative Solitons}
\end{center}
\centerline{{\large A.Solovyov}\footnote{
Kiev University, Kiev, Ukraine\\
e-mail: solovyov@gate.itep.ru\\
permanent address: solovyov@phys.univ.kiev.ua}}

\bigskip

\abstract{\footnotesize
We consider 2+1-dimensional classical noncommutative scalar field theory. The
general ansatz for a radially symmetric solution is obtained. Some exact
solutions are presented. Their possible physical meaning is discussed. The
case of the finite $\theta$ is discussed qualitatively and illustrated by
some numerical results.}

\begin{center}
\rule{5cm}{1pt}
\end{center}

\bigskip
\setcounter{footnote}{0}
\renewcommand\thefootnote{\arabic{footnote}}
\section{Introduction}
Noncommutative field theories became an object of
intense study during the last year due to their close connection to string
theory.  Namely, in the large B-field limit the associative string field
algebra $\cal A$ factors into an algebra that acts on the string center of
mass and an algebra that describes all other degrees of freedom[1,2,3]. The
second algebra seems to be terribly complicated while the first one leads to
the effective noncommutative theory. Specifically, description of the scalar
field theory of tachyons on the world-volume of unstable Dp-branes in
superstring theory is a field of employment for noncommutative theory[4].

Of especial interest are solutions of the
equation $\phi\star\phi=\phi$, i.e.  projection operators.  Here $\star$
denotes the nonlocal (Moyal) star product that becomes \be f\star g(x)=
e^{\frac{i}{2}\epsilon^{\mu\nu}\partial_{\mu}^{'}\partial_{\nu}^{''}}
f(x')g(x'')|_{x'=x''=x}
\ee
when written in rescaled coordinates.
If one goes to the momentum space,
\be
\tilde{f}(k)=\int{\frac{d^{2}x}{2\pi}f(x)e^{-ikx}},
\ee
then the $\star$-product takes the form
\be
\widetilde{f\star g}(p)=\int{\frac{d^{2}k}{2\pi}\tilde{f}(k)\tilde{g}(p-k)
e^{-\frac{i}{2}\epsilon^{\mu\nu}k_{\mu}p_{\nu}}}.
\ee
After rescaling the coordinates the (static) energy functional is
given by \be E[\phi]=\frac{1}{g^2}\int{d^{2}x (\frac{1}{2}\partial_{\mu}\phi
\partial^{\mu}\phi+\theta V(\phi)}).  \ee Here
$\frac{\partial}{\partial\phi}V(\phi)|_{\phi=0}=0$ is implied, $\theta$ is
the noncommutativity parameter[5].  All the fields are multiplied using the
$\star$-product.

The trick used to solve the equations of motion is to formally identify the
noncommutative quantum space on which the $\phi$-field lives with the
classical phase space of a single particle and to recast the problem into the
operator formalism using the Weyl-Moyal (WM) correspondence. The latter is the
correspondence between the multiplication (composition) of operators acting
on a single-particle Hilbert space and the $\star$-product of functions on
the corresponding phase space. It is defined by the commutativity of the
following diagram:

\[\begin{array}{ccc}
 \hat{f}, \hat{g}   &   \longrightarrow   &   \hat{f}\hat{g}    \\
 \Big\downarrow\vcenter{%
  \rlap{$\scriptstyle{\mathrm{\Omega^{-1}}}$}}
          &    &
 \Big\downarrow\vcenter{%
  \rlap{$\scriptstyle{\mathrm{\Omega^{-1}}}$}}    \\
 f, g   &   \longrightarrow   &   f\star g
 \end{array}                                  \]
$\Omega$ denotes a map (more precisely, one-one correspondence) from the
algebra of the functions to the algebra of corresponding operators using
symmetric (Weyl) ordering prescription:
\be
\Omega:f(p,q)\rightarrow \hat{f}(\hat{p},\hat{q})=\int{\frac{d^2k}{2\pi}
\tilde{f}(k_{p},k_{q})e^{i(k_{p}\hat{p}+k_{q}\hat{q})}},
\ee
$\tilde{f}(k)$ being defined by (2). Thus the
$\star$-product
\footnote{Here the same as in (1),(3). In our considerations $\hbar=1.$}
provides such a deformation of the
algebra of functions on the phase space (which is restricted to be a
symplectic manifold with a constant Poisson bi-vector $\theta^{\mu \nu}$)
that $\Omega$ becomes an isomorphism[6].  The following relation also
holds:  \be \int{\frac{d^{2}x}{2\pi}\phi(x)}=Tr_{\cal H}\hat{\phi}.  \ee If
the Hilbert space $\cal H$ is the space of wave functions in coordinate
representation, there exists the explicit formula to map a kernel of a
Hilbert-Schmidt operator acting on $\cal H$ into the corresponding function
\be
f(p,q)=\int{\int{dx dy K(x,y)e^{-ip(x-y)}\delta(q-\frac{x+y}{2})}}
\ee
such that $\langle x\mid\hat{f}\mid y\rangle=K(x,y)$; Weyl ordering has been
used[7].  The inverse formula reads \be K(x,y)=\int{\int{\frac{dp
dq}{2\pi}f(p,q)e^{ip(x-y)}\delta(q-\frac{x+y}{2})}} \ee and becomes a kind of
Fourier transform after integration over $q$ (in order to obtain these two
formulae one reorders $\hat{p},\hat{q}$ in the exponent in the rhs of (5)
using the Baker-Campbell-Hausdorff formula and calculates the matrix elements
directly).

The following treatment beyond the WM correspondence also exists:
one can avoid explicit mentioning the WM correspondence
at all and omit the Hilbert space of wavefunctions by using
the integral transform (7) immediately, then the $\star$-product of
two functions $f\star g$ corresponds to the convolution of their kernels
(i.e. integral transforms(8)):
\footnote{To verify this one writes the $\delta$-function in (7) in the
momentum representation
\be \delta(q-\frac{x+y}{2})=\int{\frac{d\xi}{2\pi}
e^{i\xi (q-\frac{x+y}{2})}}\ee
or, equivalently, uses (3).}
\be
K_{f\star g}(x,y)=\int{d\xi K_{f}(x,\xi)K_{g}(\xi,y)}.
\ee

In the large noncommutativity limit one can neglect the kinetic term
and obtain the following equation of motion:  $\frac{\partial V}{\partial
\phi}=0$, where $V(\phi)$ is considered a polynomial or an analytic function
(note that the constant term in the lhs is absent). Projection operators are
the basic thing to solve such an equation. A localized solution will be
referred to as a soliton.

In the next section the general form of any solution radially
symmetric w.r.t. spatial coordinates is derived. The third section deals with
some distribution-valued solutions and related topics. The fourth one is
reasoning, mainly perturbative and numeric, how does the kinetic term affect
the physics at finite $\theta$.

\section{General radially symmetric solution}

In[5] radially symmetric projection operators diagonal in the holomorphic
representation and corresponding functions on the phase space have been worked
out. The reason for the present examination is that in general adding some
non-radially symmetric functions ($\mid m\rangle\langle n\mid$'s in the
operator formalism, where $m\neq n$) or composing a series from them one can
in principle obtain a radially symmetric result (for example,
$p^{2}+q^{2}=r^{2}$). In this section we are to construct the general ansatz
for any radially symmetric solution.
In order to do this we examine how $\phi$
changes under an infinitesimal coordinate rotation $\delta q=\epsilon p$,
$\delta p=-\epsilon q$:
\be
\delta\phi(p,q)=\epsilon(p\frac{\partial\phi}{\partial q}
-q\frac{\partial\phi}{\partial p})+o(\epsilon).
\ee
(7) can be rewritten as
\be
\phi(p,q)=2\int\limits_{-\infty}^{+\infty} {d\xi K(q-\xi,q+\xi)e^{2ip\xi}}.
\ee
After use of (11) one calculates (up to the first order in $\epsilon$)
\be 0=\int\limits_{-\infty}^{+\infty} {d\xi
(p(\partial_x+\partial_y)K(x,y)-2iq\xi K(x,y))|_{x=q-\xi,y=q+\xi}
e^{2iP\xi}}.
\ee
Being multiplied by $e^{-2ip\xi'}$ and integrated w.r.t. $p$, it becomes
\be
0=\int\limits_{-\infty}^{+\infty}
{d\xi (\frac{i}{2}(\partial_x+\partial_y)K(x,y)\partial_{\xi'}-2iq\xi
K(x,y))|_{x=q-\xi,y=q+\xi}\delta(\xi-\xi')},
\ee
what simplifies to
\footnote{This equation is Fourier self-dual and remains unchanged under the
transform \be
K(x,y)=\int{\frac{d^{2}k}{2\pi}\tilde{K}(k)e^{i(k_{x}x-k_{y}y)}}.\ee}
\be
(-\frac{1}{2}\partial_{xx}+\frac{1}{2}x^{2})K(x,y) =
(-\frac{1}{2}\partial_{yy}+\frac{1}{2}y^{2})K(x,y).
\ee
Separating variables $K(x,y)=\Psi_{x}(x)\Psi_{y}(y)$ one finds \be
(-\frac{1}{2}\partial_{xx}+\frac{1}{2}x^{2})\Psi_{x}(x)=E\Psi_{x}(x),\\
(-\frac{1}{2}\partial_{yy}+\frac{1}{2}y^{2})\Psi_{y}(y)=E\Psi_{y}(y).
\ee
 We are looking for such kernels that convolution of K with K exists
(i.e. the integral in the rhs of (10) converges).  This means that $\Psi_{x}$
(and $\Psi_{y}$) vanishes sufficiently rapidly when
$x\rightarrow\infty$ (correspondingly $y\rightarrow\infty$). Similar
asymptotic conditions follow from the finiteness of energy.
\footnote{Thus the WM correspondence is valid.}
There exist some solutions for (17) only
when $E=(n+\frac{1}{2})$ and those are the wave functions of the 1d harmonic
oscillator, i.e. the basis of the holomorphic representation discussed in[5].
Thus any radially symmetric solution can be represented as
$\sum\limits_{n=0}^{\infty}{C_{n}\phi_{n}}$, where $C_{n}$ are some
coefficients and $\phi_{n}$ are solutions built in[5].

\section{On some generalized solutions}
\paragraph{Explicit construction.}
  Now we proceed to apply these considerations to the case of the SUSY even
$\phi^4$ potential
\be
V(\phi)=-\frac{1}{2}\phi^2+\frac{1}{4}\phi^4
\ee
The "improper" sign to the quadratic term can describe tachyons as their
squared mass is negative. On the other hand one can make a constant shift
and substitute $\phi\rightarrow\phi+const$ thus expanding the potential
near the minimum. In the large noncommutativity limit the equation of motion
becomes

\be -\phi+\phi^{3}=0.  \ee
It has been pointed out[8] that
$\phi(x)=\pi\delta^{2}(x)$ satisfies such an equation and[5,8]
\be
\pi\delta^{2}(x)\star\pi\delta^{2}(x)=1.
\ee
Speaking commonly, how should one
understand such a relation? Here we have $\pi\delta^{2}(x)$, a
singular distribution with a compact measure $0$ support which can be dealt
with as a weak limit of a function $\Phi(x,\epsilon)$: for instance,
\be
\Phi(x,\epsilon)=\frac{1}{2\pi\epsilon}
e^{-\frac{x^2}{2\epsilon}},\epsilon\rightarrow +0
\ee
(momentum representation with a regularization factor equal to
$e^{-\frac{\epsilon k^2}{2}}$).  Then carrying out the
$\star$-multiplication in the momentum space, one finds that divergences
(products of singular distributions and their derivatives) cancel
each other (what may look like a miracle and happens due to the nonlocality
of the $\star$-product).  Considering a function like
$\phi(x)=\pi p(x)\delta(x)$, where $p(x)$ is some polynomial, one
finds that the Fourier transform mapping into the momentum space (and,
consequently, the $\star$-multiplication) does not distinguish between
$"p(x)"$ and $"p(0)"$. This implies that any function \be \phi(x)=\pi
p(x)\delta^{2}(x) \ee such that $p(x)$ is a polynomial and \be p(0)=1 \ee
solves (19). At first glance this result seems both evident and
trouble-looking, so a little discussion on nonlocal product of distributions
seems to be necessary.

First we note that
\be
\partial_{x}^{k}(f(x)\delta(x))=f(0)\partial_{x}^{k}\delta(x)
\ee
in a weak sense (as a distribution), where $f$ is a regular function. As an
evidence for this one can integrate both parts of the above equation with an
arbitrary test function and perform integration by parts $k$ times. This
observation is to prove the dependence only on $f(0)$.

Another question is to prove the convergence and to calculate the value of
the integral in the momentum space. No uniform convergence w.r.t. the
regularization parameter (like $\epsilon$ in (21)) seems to be valid, but one
imposes another regularization in the momentum space and successfully
calculates the result of the $\star$-multiplication. This approach is the
basic one in quantum field theory and the distinction in our case is that no
renormalization is needed. That is why we say "the divergences cancel each
other". So being rigorous, the result in the momentum space also converges as
a distribution.

The third thing is the series
\be
\pi\delta^{2}(x)=\sum\limits_{n=0}^{\infty}{(-1)^{n}\phi_{n}(x)}
\ee
converges only as a distribution. For example, replacing $(-1)^{n}$ with
$(-\epsilon)^{n}$, one gets a suitable regularization. In particular the
vacuum energy becomes finite and equal to
\be
E(\epsilon)=-\frac{1+2\epsilon^{2}}{4(1-\epsilon^{4})},
\ee
but $E\rightarrow-\infty$ as $\epsilon\rightarrow 1-0$.
Evidence that the configuration $C_{n}=(-1)^{n}$ is situated at the boundary
point of the strong convergence range enables one to put forward the
equivalent mathematical approach to the polynomial $p(x)$ as a factor
changing the form of the regularization.

The above considerations can also be applied to power series with reasonable
asymptotic behaviour of the coefficients.

It is interesting to
see what is the corresponding operator in the Hilbert space. From (8) one
easily calculates $K(x,y)=\delta(x+y)$, i.e. this is the spatial reflection
$\hat{P}$ in imaginary particle's configuration space. That is the obvious
reason why it is no projector. Corresponding coefficients $C_{n}$ are
simply the matrix elements of $P$ in the holomorphic representation:
$C_{n}=(-1)^{n}$ as the harmonic oscillator wavefunctions have definite
parity. For the unity operator, of course, $C_{n}=1$ and we can construct a
projector onto even states $\frac{1}{2}(\hat{I}+\hat{P})$,
 $\phi=\frac{1}{2}(1+\pi p(x)\delta^{2}(x))$.

\paragraph{Classical stability.}
Let us examine whether our solution $C_{n}=(-1)^{n}$, $\theta=\infty$ is
stable. For this purpose we expand the energy functional near this
configuration up to the second order terms and check if the quadratic form to
appear is positively defined. No radial symmetry of the infinitesimal
field change $\delta\phi$ is implied. The second variation proves to be (in
operator formalism)
\footnote{We have performed some cyclic permutations under the trace sign.
Field variations are considered to be small.}
\be \delta E[\phi]=
\frac{2\pi\theta}{g^2}Tr_{\cal H}(-\frac{1}{2}(\widehat{\delta\phi})^{2} +
\frac{1}{4}(4\hat{\phi}^{2}\widehat{\delta\phi}^{2} +
2(\hat{\phi}\widehat{\delta\phi})^{2})),
\ee
what by use of $\hat{\phi}^{2}=\hat{I}$ reduces to
\be
\delta E[\phi]=
\frac{\pi\theta}{g^2}Tr_{\cal H}(\widehat{\delta\phi}^{2} +
(\hat{\phi}\widehat{\delta\phi})^{2})>0.
\ee
Thus the solution is stable. The same considerations apply to $\phi=1,$
 $C_{n}=1$. It is clear that any solution of (19) to satisfy
\be \phi\star\phi=1 \ee
is classically stable while the solution for which exists $n$ such that
$C_{n}=0$ is unstable, for instance, to the radial fluctuations of the $n$-th
mode because the quartic term gives no rise to the quadratic part of
expansion now.

\paragraph{Physical interpretation.}
First, "localization" of solution is understood comparatively as the energy of
the $\delta$-like solution is infinite, but it is perfectly localized
spatially.

Coefficients of the polynomial $p(x)$ are unlikely to be important
on the classical level and seem to be some internal degrees of freedom. The
question concerns their meaning.

In general there exist some different treatments with noncommutative solitons.

1. Such a localized 2d configuration can be considered to be a D23-brane
within D25 brane[8,9]. The correct brane tension is recovered from the action
calculated on the noncommutative soliton. $P(x)$ can describe excitations
living on the world volume of this D23-brane.

2. In[8] it was proposed that noncommutative solitons are the classical vacua
for the compact scalar field theory. In our case this statement is supported
by the fact that \be
E[\phi]|_{\phi=\pi\delta^{2}(x)}=E[\phi]|_{\phi=1}=-\infty,
\ee
while the energy density remains constant and equal to
\be
\varrho_{E}=-\frac{\theta}{4g^{2}}.
\ee
The energy density does not depend on the sign of $C_{n}$ if $C_{n}=\pm1$ at
all as $\phi_{n}$ are $\star$-orthogonal. Then such classical vacua
can be parametrized by the infinite
number of coefficients $C_{n}=\pm 1$. The coefficients of the polynomial
$p(x)$ are some excitations. Similar considerations concerning the energy
density partially apply to the next item.

Following these ideas, it is interesting to exhibit
an example of an instanton interpolating between such vacua. To do this
we first consider the scalar self-interacting field theory in
2+1-dimensional Minkowski space-time with the $(+--)$ signature and
commutative timelike direction. The action functional is
\footnote{No rescaling has been implied yet.}
\be
S[\phi]=\frac{1}{g^2}\int{d^{3}x
(\frac{1}{2}\partial_{\mu}\phi\partial^{\mu}\phi - V(\phi)), }
\ee
from what one reads the energy functional (4) and (now in rescaled
noncommuting space coordinates) the non-static equation of motion
\be
\frac{\partial^{2}\phi}{\partial t^{2}}=\frac{1}{\theta}\bigtriangleup\phi
-\frac{\partial V}{\partial\phi}.
\ee
When $\theta=\infty$ the first term in the rhs can be neglected.
In the euclidean time (or, equivalently, inverted potential)
\be
\frac{\partial^{2}\phi}{\partial t^{2}}= \frac{\partial V}{\partial\phi}.
\ee
Now it is easy to construct an example of an instanton tunneling between
two noncommutative vacua. Should one restrict the solution to be diagonal at
any time moment $-\infty<t<\infty,$ the equation of motion becomes the
infinite set of decoupled equations
\be
\frac{d^{2}}{dt^{2}}C_{n}=C_{n}-C_{n}^{3},
\ee
each of them describes the well known quantum mechanical double well kink if
the coefficient $C_{n}$ is tunneling from -1 to 1, antikink ($1\rightarrow
-1$) or a constant ($1\rightarrow 1$ or $-1\rightarrow-1$, no tunneling).

This construction can be applied to the dynamics of radially symmetric
solitons and the equations of motion are written in the Minkowski space-time.
Recently some new results concerning noncommutative soliton scattering[10]
were obtained using the adiabatic approximation.
\footnote{The calculations below are closed for $\theta=\infty$.}
The typical feature is that
the equations (35) are identical for all modes. If all the initial conditions
are the same (up to a sign), the solution is $\phi(r,t)=C(t)\Phi(r)$, i.e.
the time dynamics reduces to the scaling factor in front of the soliton.  As
$V(\phi)\rightarrow\infty$ when $\phi\rightarrow\infty$, all the coefficients
$C_{n}$ remain finite.

If small excitations mentioned above are radially symmetric, they can be
described as small oscillations of $C_{n}$ near the minimum of potential.

3. Some vortex-like solutions in noncommutative abelian Higgs model (which
has similar potential for the scalar field) have recently been discussed[11].
In this context (and also for the brane interpretation) the subtlety to arise
is the uniform energy density while it is significantly singular for a
commutative brane or a vortex. It is not completely clear for what vortex's
internal degrees of freedom $p(x)$ is responsible.

\section{Behaviour of solutions with finite $\theta$}
\paragraph{Recurrent relation.}
The question still unanswered completely is how does the kinetic term affect
the physics at finite $\theta$. To examine some aspects one should write the
equations of motion at finite $\theta$. It is known[5,12] that in the operator
formalism
\be
\frac{1}{2}\partial_{\mu}\phi\star\partial^{\mu}\phi \leftrightarrow
[\hat{a},\hat{\phi}][\hat{\phi},\hat{a}^{\dagger}],
\ee
where $\hat{a}^{\dagger},$ $\hat{a}$ are creation and annihilation operators
defined in Appendix. Performing variation in (4) w.r.t. $\phi$, one obtains
\be
[\hat{a}^{\dagger},[\hat{a},\hat{\phi}]] +
[[\hat{\phi},\hat{a}^{\dagger}],\hat{a}] = - \theta \frac{\partial V}{\partial
\phi}(\hat{\phi}).
\ee
By use of Jacobi identity and commutation relations for creation and
annihilation operators lhs yields
\be
[\hat{a}^{\dagger},[\hat{a},\hat{\phi}]] = -\frac{1}{2} \theta \frac{\partial
V}{\partial \phi}(\phi).
\ee
With the radially symmetric ansatz the equation of
motion simplifies to the recurrent relation
\footnote{For derivation see Appendix.}
\be
(n+1)C_{n+1}-(2n+1)C_{n}+nC_{n-1}=\frac{1}{2}\theta V^{'}(C_{n}), n>0, \\
C_{1}-C_{0}=\frac{1}{2}\theta V^{'}(C_{0}), n=0.
\ee
Should one try to find an appropriate potential $V(\phi)$ for a solution
$C_{n}=(-1)^{n}A$ to exist, it will end in no positive result as in this case
(39) reduces to
\be
2(-1)^{n+1}(2n+1)A=\frac{\theta}{2}V^{'}((-1)^{n}A).
\ee
As it has been emphasized, an equation like (39) is hard to solve, but some
effects could be studied qualitatively or numerically.

\paragraph{Qualitative effects.}

1.The differential equation for this scheme is
\be
nC^{''}+C^{'}=\frac{1}{2}\theta C(C^{2}-1).
\ee
A similar one occurs in the construction of the usual commutative
Abrikosov-Nielsen-Olesen vortex and the only acceptable asymptotic for us is
that the coefficients $C(n)=C_{n}$ vanish at the infinity. The mode expansion
$\phi=\sum\nolimits{C_n}\phi_{n}$ is somewhat like a discrete Fourier
transform.
Then the singular (or at least localized) vortex-like solution in the
discrete momentum space should lead to delocalization in the field
$\phi$-space. Thus, the $\delta$-like
solution becomes finite-size. The effect of finite $\theta$ is imposing a
regularization.

2.The energy functional is[12]
\be
E[\phi]=\frac{1}{g^{2}}
\sum\limits_{n=0}^{\infty}{((2n+1)C_{n}^{2}-2(n+1)C_{n+1}C_{n} +
\theta V(C_{n}))}.
\ee
Analyzing the first two terms as a perturbation, i.e. simply calculating
the energy functional for the solution at $\theta=\infty$ one observes that
the second addend in the kinetic term causes the level splitting and breaks
the $2^{\infty}$-times degeneration.  Perfectly localized
$\delta$-like solution becomes the highest energy level because all the
perturbative terms in (41) appear to be positive for the coefficients with
alternating sign. This is the reason to understand of what importance is the
$\delta$-like radially symmetric solution. To do that one observes that the
results for $\theta=\infty$ are true for any set of mutually orthogonal
projectors. It is the kinetic term that causes the $U(\infty)$ symmetry
breakdown. For an arbitrary operator like
\be
\hat{\phi}=U^{\dagger}diag(\{\lambda_{n}\})U,
\ee
with $U$ being unitary the energy functional is equal[12] up to the coupling
constant
\be
E[\{\lambda_{n}\},\{U_{mn}\}]=
\sum\limits_{n=0}^{\infty}{\lambda_{n}^{2}(1+
2\sum\limits_{m=0}^{\infty}{m|U_{mn}|^{2}})}
-2\sum\limits_{m,n=0}^{\infty}{\lambda_{m}\lambda_{n}|A_{mn}|^{2}}
+\theta\sum\limits_{n=0}^{\infty}{V(\lambda_{n})},
\ee
where
\be
A_{mn}=\sum\limits_{k=1}^{\infty}{\sqrt{k}U_{kn}U_{k-1,m}^{*}},
\ee
If we hold $\{\lambda_{n}\}$ fixed, then the first order variation of $E$
w.r.t.  $U$ is
\be \
\delta E=\frac{\partial E}{\partial
U_{mn}}|_{U=1}\delta U_{mn} + \frac{\partial E}{\partial
\bar{U}_{mn}}|_{U=1}\delta \bar{U}_{mn} =\\
2\sum\limits_{n=0}^{\infty}{(3n+1)(\delta U_{nn}+\delta\bar{U}_{nn})}=0
\ee
as in the first order
\be
\delta U_{nn}+\delta\bar{U}_{nn}=0
\ee
for $U$ is unitary.

Thus for $\theta=\infty$ the matrix $U$ serves as coordinates along the
vacuum valley, but at finite $\theta$ the special significance of radially
symmetric solutions becomes clear.

\paragraph{Numeric study.}
Our method is different from that in[12].
It is clear that specifying $C_{0}$ defines all the other coefficients
and provides an opportunity for the numeric study.
In our study $\theta=100$ and remains constant. The main goal is to see how
does a finite-$\theta$ solution look like.
Some graphs are presented below for $C_{0}$ = $10^{-10}$.

\begin{figure}[p]
\epsffile{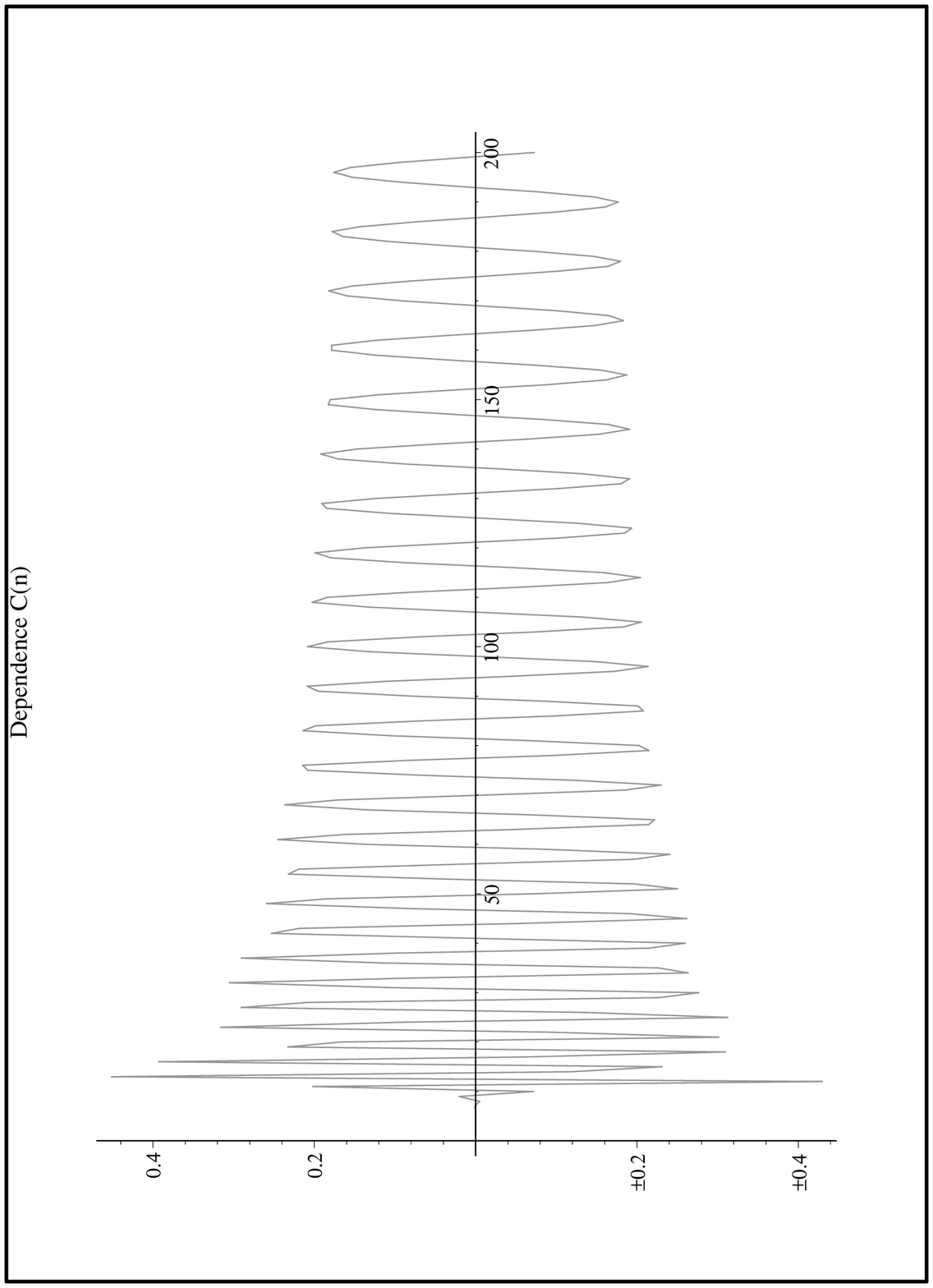}
\end{figure}
\begin{figure}[p]
\epsffile{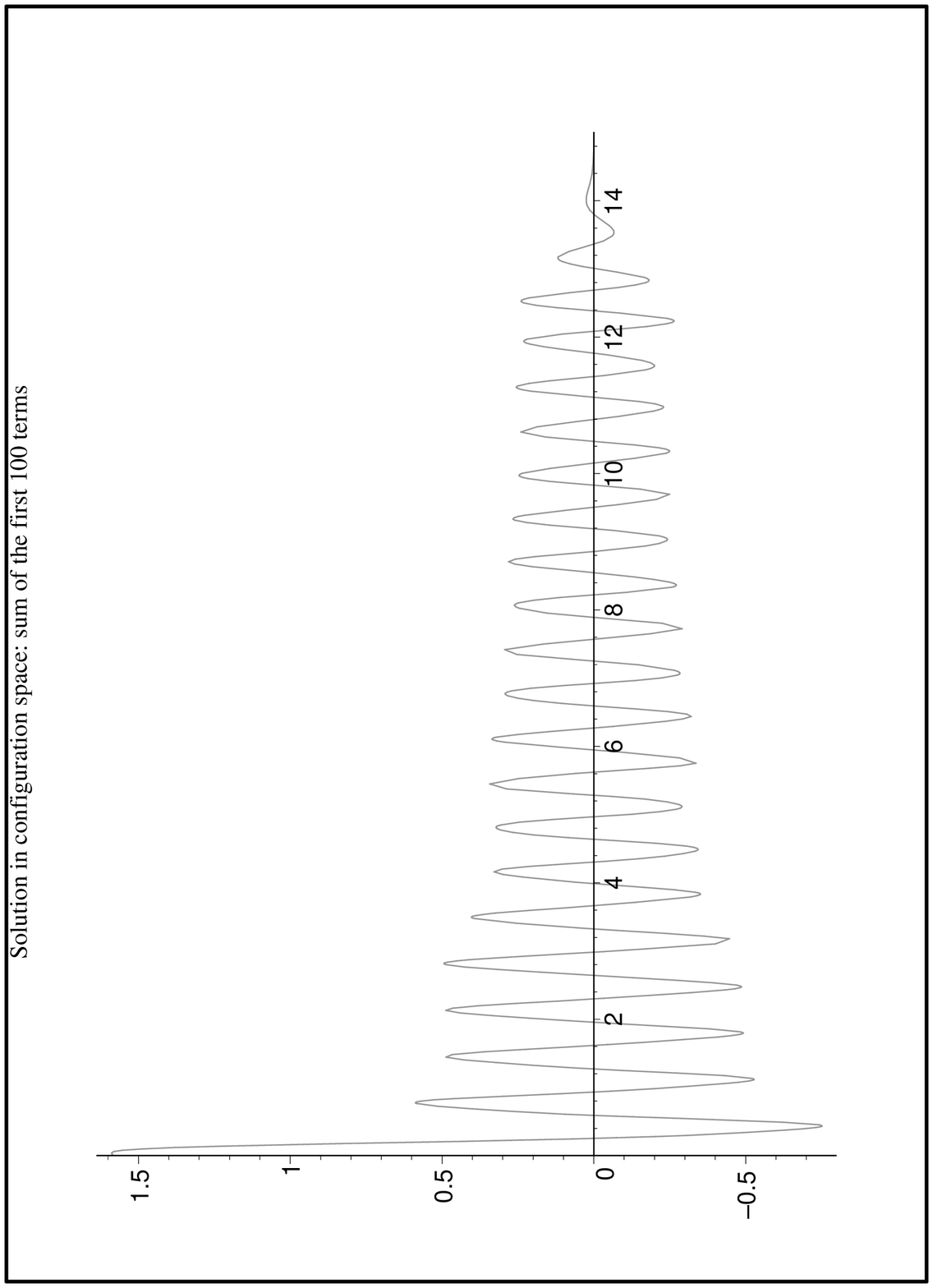}
\end{figure}
\begin{figure}[p]
\epsffile{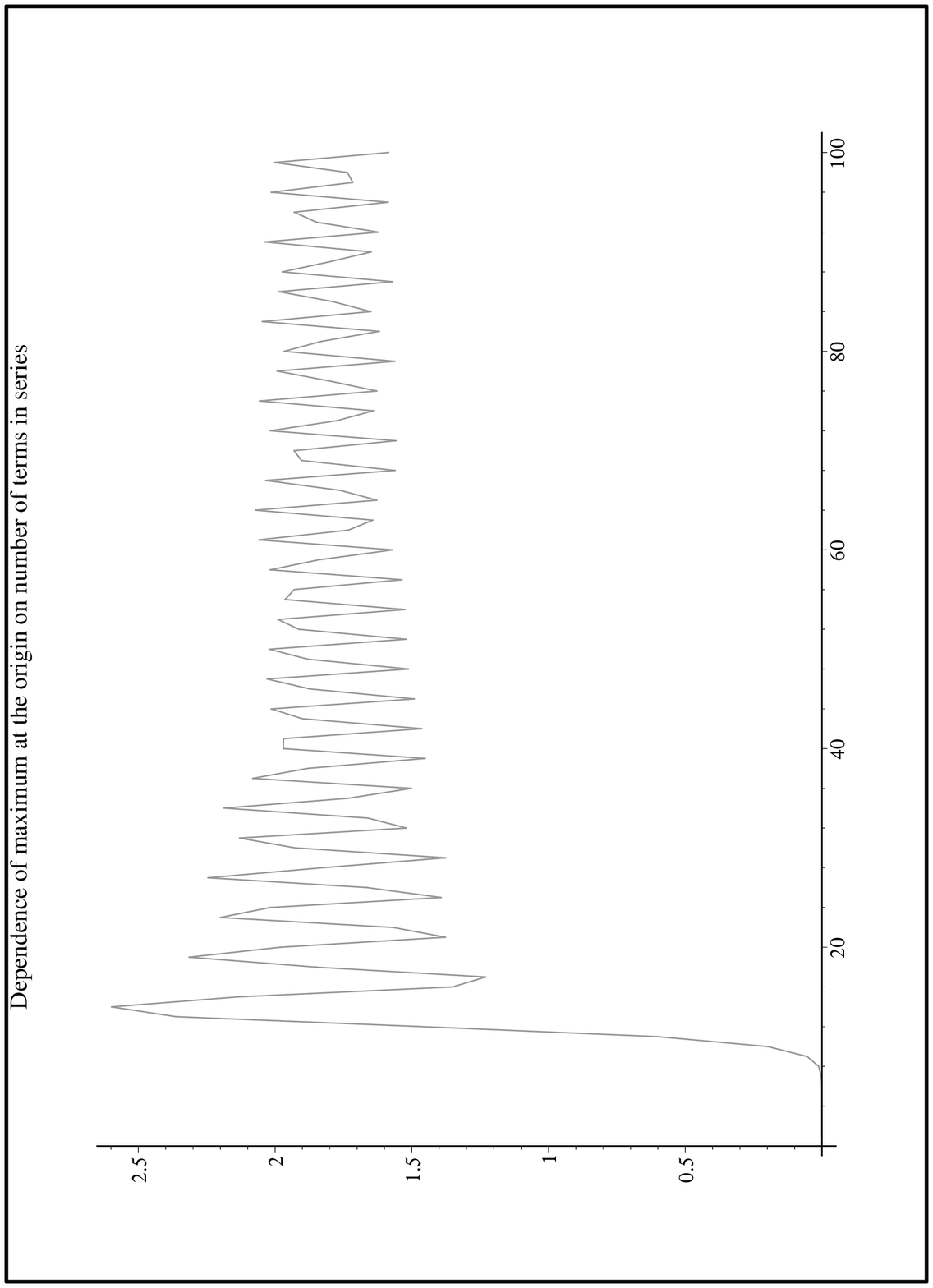}
\end{figure}
The equations are strongly nonlinear, so as we start from $C_{0}=10^{-10}$ it
first grows in ten orders up to $C \sim 1$ and then slowly converges to $0$
(see the first graph). In the second graph we show the sum of the 100 first
terms in the series for $\phi(r)$. The third one illustrates that $\phi(0)$
remains finite and thus the solution is finite-size.

Another interesting phenomenon is that a small change in $C_{0}$ can make
the solution divergent. For example, there exist at least two
disconnected ranges of convergence $C_{0} \leq 3.82\cdot 10^{-10}$ and
$9.1\cdot 10^{-10}\leq C_{0} <11.6\cdot 10^{-10}$, $\theta=100$. This point
agrees well with the soliton hierarchy first observed in[12].

\section{Concluding remarks}
In the present paper we consider 2+1-dimensional noncommutative
classical scalar field theory. The standard and basic tool for solving
equations of motion at $\theta=\infty$ is the Weyl-Moyal correspondence
between the product of operators acting on the Hilbert space and the nonlocal
product of functions. We mention how one can solve the equations of motion in
principle without referring to the Weyl-Moyal correspondence. Then the
general ansatz for a radially symmetric solution is obtained confirming the
celebrated results of[5]. Next these results are applied to studying
generalized (distribution-valued) solutions and their physical assignment. The
$\star$-product of distributions is defined as a distribution in the momentum
space. Small excitations are shown to change the regularization. An approach
to the accurate study of the radially symmetric solution dynamics is given
and the effect of the time dynamics coming to the field rescaling is
described. Then we proceed to examine the physics at finite $\theta$. The
effects of delocalization and taking down the degeneration are analyzed
qualitatively and numerically. The special role of radially symmetric
solutions is clarified.

There arise different questions of interest such as dynamics of fields
without radial symmetry and general scattering problem. It is
interesting to take quantum corrections into account. Among other problems
there are coupling with the gauge field
and searching out some closed form solution and analysis for finite $\theta$.

\section{Acknowledgements}
Author is greatful to ITEP group for hospitality and useful discussions (in
particular to S.Gukov, K.Selivanov, Yu.Sitenko, V.Shadura, A.Zotov, A.Chervov, A.Alexandrov,
D.Melnikov, S.Klevtsov, A.Dymarsky) and especially to
A.Gorsky and A.Morozov for introducing the problem and discussions.

\section{Appendix}

\paragraph{Derivation of the recurrent relations.}
In the holomorphic representation the key role is played by the two operators
\be
\hat{a}=\frac{\hat{q}-i\hat{p}}{\sqrt{2}},  \\
\hat{a}^{\dagger}=\frac{\hat{q}+i\hat{p}}{\sqrt{2}}.
\ee
Commutation relations for them $[\hat{a},\hat{a}^{\dagger}]=\hat{I}$ do have a
representation in the space of analytic functions $f(\bar{z})$ with the inner
product[13]
\be
\langle f_{1},f_{2}\rangle=\int{\frac{dz d\bar{z}}{2\pi\i}
e^{(-\bar{z} z)}\overline{f_{1}(\bar{z})} f_{2}(\bar{z}}).
\ee
The orthonormal state vectors are represented as
\be
\mid n \rangle \leftrightarrow
\psi_{n}(\bar{z})=\frac{(\bar{z}^{n})}{\sqrt{n!}},
\ee
creation and annihilation operators as
\be
\hat{a}^{\dagger} \leftrightarrow \bar{z}, \\
\hat{a} \leftrightarrow \bar{\partial}.
\ee
An arbitrary operator $\hat{U}$ with matrix elements equal to
\be
U_{mn}=\langle \psi_{m} \mid \hat{U} \mid \psi_{n} \rangle
\ee
can be represented as the integral operator with the kernel
\footnote{$A(z,\bar{w})$ is an analytic function (some
series) of two complex arguments, and in general $z \neq w$.}
\be
A(\bar{z},z)=\sum\limits_{m,n} {U_{mn}
\frac{\bar{z}^{m}}{\sqrt{m!}} \frac{z^{n}}{\sqrt{n!}}}, \ee \be
(\hat{U}f)(\bar{z})=\int{\frac{dz d\bar{z}}{2\pi\i}e^{(-\bar{z} z)}
A(\bar{z},z)f(\bar{z})}.
\ee
Now all the needed commutators in the equation of motion can be simply
computed using the formulae
\footnote{The last two ones are obtained using integration by parts and
reflect the fact that $\hat{a}$, $\hat{a}^{\dagger}$ are hermitian
conjugate.}
\be
\hat{a}^{\dagger}\hat{U} \leftrightarrow
\bar{z} A(\bar{z},z), \\ \hat{a}\hat{U} \leftrightarrow \bar{\partial}
A(\bar{z},z), \\ \hat{U}\hat{a}^{\dagger} \leftrightarrow \partial
A(\bar{z},z), \\ \hat{U}\hat{a} \leftrightarrow z A(\bar{z},z).
\ee
Under the assumption of the radial symmetry $U_{mn}=C_{m}\delta_{mn}$ the
result is
\be
( \partial\bar{\partial} - z\partial - \bar{z}\bar{\partial} + z\bar{z} -1 )
A(\bar{z},z) = \frac{\theta}{2}\sum\limits_{n=0}^{\infty}\frac{\partial
V}{\partial \phi}(C_{n}) \psi_{n}(\bar{z})\overline{\psi_{n}(\bar{z})},
\ee
and after the
simple direct calculation of matrix elements one arrives to (39).
The energy functional is obtained in a similar way.

\end{document}